\documentstyle[12pt]{article}

\def\sp{{\hbox{\rm Spin}}}

\def\be{\begin{equation}}
\def\ee{\end{equation}}
\def\bea{\begin{eqnarray}}
\def\eea{\end{eqnarray}}

\begin{document}
\begin{titlepage}
\begin{flushright}
{ ~}
US-FT-5/97\\
hep-th/9702054
\end{flushright}
\vglue 1cm
\begin{center}
{\LARGE Twisted Baryon Number in $N=2$ Supersymmetric QCD}
\vskip3cm

{\large J. M. F. Labastida\footnotemark 
\footnotetext{e-mail: labastida@gaes.usc.es}
and M. Mari\~no\footnotemark \footnotetext{e-mail: marinho@gaes.usc.es} }

\vspace{1cm}

{\it  Departamento de F\'\i sica de
Part\'\i culas,\\ Universidade de Santiago
de Compostela,\\ E-15706 Santiago de Compostela, Spain.}

\vskip2cm

{\large \bf Abstract}
\end{center}
We show that the baryon number of $N=2$ supersymmetric QCD can be twisted in
order to couple the  topological field theory of non-abelian monopoles to
$\sp^c$-structures.  To motivate the construction, we also consider
some aspects of the twisting  procedure as a gauging of  global currents in
two and four dimensions, in particular the r\^ole played by anomalies.

\vfill

\end{titlepage}

\def\theequation{\thesection.\arabic{equation}}

\section{Introduction}

One of the most impressive consequences of the work by Seiberg and Witten
\cite{swone, swtwo}  has been  the dual description of Donaldson theory in
terms of monopole equations \cite{mfm} (see  \cite{marcolli, morgan, gptres,
tribu, don} for a review). This has opened the way  to new developments  in
four-dimensional topology and has provided astonishing new links between this 
field and  the dynamics of supersymmetric gauge theories in four dimensions.
One of these developments  is the non-abelian generalization of the abelian
Seiberg-Witten equations, known as  non-abelian monopoles, which have been
extensively studied both from the mathematical  \cite{okone, 
teleman, pt, bradgp}  and the physical point of view \cite{lmna, lmpol, lmt,
tqcd} in the last two years. 

It is well-known that the Seiberg-Witten monopole equations involve 
$\sp^c$-structures on  a four-manifold \cite{sping}, and it was shown in
\cite{abelia} that these  naturally arise when  considering the duality
transformations of the underlying physical theory.  For the non-abelian case,
$\sp^c$-structures were considered in \cite{tqcd} in order to  define twisted
$N=2$ supersymmetric QCD on a general four-manifold. In this note, we will
propose a  different way to couple matter hypermultiplets to
$\sp^c$-structures in non-abelian  monopole theories. The idea is to consider
an extended twisting procedure by  gauging an additional global symmetry of
the  physical theory. The global symmetry turns out to be the baryon number
$U(1)_B$. We will  show in detail that the procedure is consistent and that
the quantum numbers associated to  this symmetry are the appropriate ones
from the geometrical point of view. To motivate  the construction, we will
review some aspects of the twisting as a gauging procedure.  A general
conclusion of our analysis is that global symmetries not necessarily 
associated to the supersymmetry algebra could be considered in the
construction of  a topological field theory.   

The organization of this paper is as follows: In section two we consider the
twisting as a gauging in Donaldson-Witten theory and  in two-dimensional
topological sigma models. In section three we consider the  gauging of the
$U(1)_B$ current in $N=2$ supersymmetric QCD and the coupling to
$\sp^c$-structures.  Finally, in section four we present our conclusions and
outlook.

\section{The twist as a gauging}
\setcounter{equation}{0}

Usually, the twisting procedure has been understood as a modification of the 
rotation group of a supersymmetric $N=2$ theory through an embedding of the
global  symmetry associated to the supersymmetry algebra \cite{tqft, tsm, lla,
altwo}.  An alternative (and equivalent) point of view is obtained when the
twisting  procedure is regarded as a gauging of an internal symmetry group, in
which a  global symmetry of the underlying supersymmetric model is promoted to
a  space-time symmetry. In many cases, the gauging is performed by adding to 
the Lagrangian of  the original theory a new term, involving the coupling of
the internal current  to the Spin connection of the underlying manifold
\cite{egu,ceco,cmr}. 

We will now discuss this approach to the twisting procedure in some detail in
the  case of $N=2$ supersymmetric Yang-Mills, the model originally considered
by Witten in
\cite{tqft}. The field content of the minimal $N=2$ supersymmetric 
Yang-Mills theory with gauge group $G$ on ${\bf R}^4$ is the following: a
gauge field $A_{\mu}$,   two Majorana spinors
$\lambda_{i\alpha}$, $i=1,2$, and their conjugates 
$\overline\lambda^{i\dot\alpha }$, a complex scalar $B$, and an auxiliary
field $D_{ij}$ (symmetric in $i$ and $j$). The indices $i$, $j$ denote  the
isospin indices of the  internal symmetry group $SU(2)_I$ of $N=2$
supersymmetry. The two Majorana spinors  $\lambda_1$, $\lambda_2$ form a
doublet of $SU(2)_I$. All these fields are considered in the adjoint
representation of the gauge group $G$. The $SU(2)_I$ current of this model is
given by: 
\be
j^{\mu}_a={\overline \lambda}{\sigma}^{\mu}{\sigma_a}\lambda,
\label{corriente}
\ee
where ${\sigma}^{\mu}=({\bf 1}, i\sigma_a)$, and $\sigma_a$ are the Pauli 
matrices. 

If we try to formulate the theory on a general four-manifold $X$ with 
Euclidean signature, the Majorana spinors $\lambda_{i \alpha}$ and 
their conjugates $\overline\lambda^{i\dot\alpha }$ are taken as 
independent Weyl spinors with opposite chirality, and we use the
prescription  of minimal coupling to gravity in the Lagrangian. The structure
group of the  spinor bundle is,  
\be
\sp_4 \simeq SU(2)_{L}\times SU(2)_{R},
\label{estr}
\ee
 so that the covariant derivative 
acting on positive chirality spinors is given by,
\be
D_{\mu}M_{\alpha}=\nabla_{\mu}
M_{\alpha}-i\omega_{\mu}^a(\sigma_{a})_{\alpha}{}^{\beta}
M_{\beta},
\label{cova}
\ee
where $\omega_{\mu}^a$ is the $SU(2)_{L}$ Spin connection on $X$, and
$\nabla_{\mu}$ is  the covariant derivative on flat space (including the gauge
connection). This construction  can always  be done  locally, but globally we
have of course the usual topological obstruction associated  to the second
Stiefel-Whitney class, $w_2(X)$. If $X$ is not Spin, we cannot consistently 
couple the original $N=2$ theory to gravity. 

The twist of the theory under consideration involves
the  gauging of the $SU(2)_I$ group as the space-time symmetry group
$SU(2)_{L}$. As the  only fields charged with respect to $SU(2)_I$ are
spinors, the gauging is achieved  after adding to the Lagrangian the coupling
of the $SU(2)_L$ connection to the  $SU(2)_I$ current:
\be
-\omega_{\mu}^a j^{\mu}_a=
-{\overline \lambda}^{i}_{\dot \alpha}(\sigma^{\mu})^{\alpha 
\dot \alpha}(\sigma_a)_i{}^j\omega_{\mu}^a \lambda_{j \alpha}.
\label{acoplo}
\ee
The only change in the Lagrangian is in the fermion kinetic term, where 
the coupling to gravity becomes,
\be
-{\overline \lambda}^{i}_{\dot \alpha}(\sigma^{\mu})^{\alpha 
\dot \alpha}\{ \delta_{\alpha}{}^{\beta}\delta_{i}{}^{j}-i\omega_{\mu}^a 
\big((\sigma_a)_{\alpha}{}^{\beta}\delta_{i}{}^{j} +
(\sigma_a)_i{}^{j}\delta_{\alpha}{}^{\beta}\big)\}\lambda_{j\beta}.
\label{kin}
\ee
The connection appearing here is the tensor product connection on the bundle 
$S^{+}\otimes S^{+}$, which is isomorphic to $\Omega^{0}_{\bf C}\oplus 
\Omega^{2,+}_{\bf C}$. The scalar part corresponds to the antisymmetric part
of  $\lambda_{w\beta}$ while the self-dual two-form corresponds to the
symmetric one.  We can write (\ref{kin}) in terms of space-time fields,
and to do this we introduce a  scalar $\eta$, a one-form $\psi_{\mu}$, and a
self-dual two-form $\chi_{\mu \nu}$ as: 
$$
\eta=-\lambda_{\beta}^{\beta},\,\,\,\,\,\,\ 
\psi_{\mu}=i(\sigma_{\mu})^{\dot \alpha \alpha} 
{\overline \lambda}_{\alpha \dot \alpha},
$$
\be
\lambda_{(w\beta)}=-{1 \over {2 {\sqrt 2}}} C^{\dot \alpha \dot \beta} 
(\bar \sigma_{\mu})_{w \dot \alpha}(\bar \sigma_{\nu})_{\beta \dot \beta}
\chi_{\mu \nu},
\label{formas}
\ee
being ${\bar\sigma}^{\mu}=({\bf 1}, -i\sigma_a)$ on flat space 
(on a curved space, the 
usual vierbein is included). 
In terms of these fields the fermion kinetic 
terms can be written, after a lengthy 
computation, as
\be
-{i \over 2}\psi^{\mu}D_{\mu} \eta - {\sqrt 2}
\psi_{\mu}D_{\nu}\chi^{\nu \mu},
\label{kindos}
\ee
which are the standard fermion kinetic terms in Donaldson-Witten theory. 
Notice that the gauging of the $SU(2)_I$ global symmetry allows
one to define the  theory on an arbitrary smooth four-manifold, as the 
spinors become the differential forms in (\ref{formas}). Although these 
differential forms are complex, they have of course a
natural underlying real structure.  One must restrict the resulting  fields to
be {\it real} differential forms, in order to have the same number of  degrees
of freedom in the untwisted and twisted theory. This  counting of degrees of
freedom must be taken into account if one is interested in  extracting some
information from the dynamics of the untwisted, physical theory. 

In general, the gauging of a global symmetry can generate anomalies in the
resulting  theory. In the case of $N=2$ supersymmetric Yang-Mills theory, the
possible  anomalies are related to  the global $SU(2)$ anomaly discovered by
Witten in \cite{sudosano}, and it is easy  to see \cite{tqft} that they only
appear when the corresponding moduli space is not  orientable. This is not
the case for the moduli space  of ASD connections \cite{dk}, which is the one
described by Donaldson-Witten theory,   and the twisted theory is then
anomaly-free. 

In the case of $N=2$ supersymmetric Yang-Mills theory in four dimensions the
twisting  involves the gauging of a $SU(2)$ internal symmetry group. In the
twisting of $N=2$  supersymmetric sigma models in two dimensions, the rotation
group is abelian, and the  twisting involves now the global $SO(2)=U(1)$
rotation group. Because of that,  the two-dimensional  situation has a formal
analogy with the twisting of the $U(1)_B$ current to be discussed  in the next
section, and we will examinate it now in some detail. 

From the point of view of the gauging procedure, it is better to start with an
$N=2$  supersymmetric sigma model involving chiral multiplets. The target
manifold  $M$ is then K\"ahler and we have a bosonic field $\phi: \Sigma
\longrightarrow M$  and a Dirac spinor $\psi^{I}_{\pm} \in  \Gamma(\Sigma,
S^{\pm}\otimes \phi^{*}(TM))$, where $TM$ is the holomorphic tangent bundle 
to $M$. The kinetic fermion term in the action is: 
\be
S_{\rm f}=\int_{\Sigma}d^{2}z ig_{{\bar I}J}
(\psi^{\bar I}_{+}D_{\bar z}\psi^{J}_{+}+
\psi^{\bar I}_{-}D_{ z}\psi^{J}_{-}),
\label{fermiondos}
\ee
where the covariant derivatives are given in local coordinates by,
\bea
D_{\bar z} \psi^{J}_{+}&=&\partial_{\bar z}\psi^{J}_{+}+
{i \over 2}\omega_{\bar z}\psi^{J}_{+}+
\Gamma^{J}_{KL}
\partial_{\bar z} \phi^{K} \psi^{L}_{+},
\nonumber\\
D_{z} \psi^{J}_{-}&=&\partial_{z}\psi^{J}_{-}
-{i \over 2}\omega_{z}\psi^{J}_{-}+
\Gamma^{J}_{KL}
\partial_{z} \phi^{K} \psi^{L}_{-}.
\label{covados}
\eea
This theory has a conserved, non-anomalous vector 
current $j^{\mu}_{V}=g_{{\bar I}J}
{\overline \psi}^{\bar I}
{\gamma}^{\mu} \psi^{J}$, with components,
\be
j^{z}_{V}=2g_{{\bar I}J}\psi^{\bar I}_{-} 
\psi^{J}_{-},\,\,\,\,\,\ j^{\bar z}_{V}=
2g_{{\bar I}J}\psi^{\bar I}_{+} \psi^{J}_{+},
\label{jvector}
\ee
and an anomalous axial current 
$j^{\mu}_{5}=g_{{\bar I}J}{\overline \psi}^{\bar I}
{\gamma}^{\mu} \gamma_5 \psi^{J}$ with components,
\be
j^{z}_{5}=2g_{{\bar I}J}\psi^{\bar I}_{-} 
\psi^{J}_{-},\,\,\,\,\,\ j^{\bar z}_{5}=
-2g_{{\bar I}J}\psi^{\bar I}_{+} \psi^{J}_{+}.
\label{axial}
\ee
The anomaly is given by the index of the Dirac operator and reads:
\be
\int_{\Sigma} \phi^{*}(c_1(M)).
\label{anomalia}
\ee 
To twist the model we can gauge the $U(1)_V$ or the $U(1)_A$ symmetries.  The
first choice  leads to the A model and the second one to the B model
\cite{tsm,lla, mm}.  As in Donaldson-Witten theory, we promote the abelian
global symmetry to a  worldsheet space-time  symmetry, and in this case this
amounts to add to the Lagrangian the coupling of the  corresponding currents
to the worldsheet Spin connection. For the A model we have, 
\bea
S_{\rm f}&-&{i \over 4} \int_{\Sigma} d^2 z {\sqrt h}\omega_{\mu}j^{\mu}_{V} 
\nonumber\\ 
&=& \int_{\Sigma} d^2 z {\sqrt h} i g_{{\bar I}J}\{ \psi^{\bar I}_{+} 
({\overline \partial} \psi^{J}_{+}+\Gamma^{J}_{KL}
{\overline \partial} \phi^{K} \psi^{L}_{+}) 
\nonumber\\ 
& &\,\,\,\,\,\,\,\,\,\,\,\,\,\,\,
+\psi^{\bar I}_{-}(\partial \psi^{J}_{-}
-i\omega_{z}\psi^{J}_{-}+ \Gamma^{J}_{KL}
\partial \phi^{K} \psi^{L}_{-}) \},
\label{tipoa}
\eea
and for the B model,
\bea
 S_{\rm f}&-&{i \over 4} \int_{\Sigma} d^2 z {\sqrt h}\omega_{\mu}j^{\mu}_{A}
\nonumber\\ 
&=&\int_{\Sigma} d^2 z{\sqrt h}ig_{{\bar I}J}\{ \psi^{\bar I}_{+} 
({\overline \partial}\psi^{J}_{+}+i\omega_{\bar z}\psi^{J}_{+}+\Gamma^{J}_{KL}
\partial_{\bar z} \phi^{K} \psi^{L}_{-})
\nonumber\\ 
& &\,\,\,\,\,\,\,\,\,\,\,\,\,\,\,
+\psi^{\bar I}_{-}(\partial\psi^{J}_{-}-
i\omega_{ z}\psi^{J}_{-}
+\Gamma^{J}_{KL}
\partial \phi^{K} \psi^{L}_{+}) \}.
\label{tipob}
\eea     
We see again that, in the twisted models, the fermion fields 
have changed their spin
content.  $\psi^{J} _{-}$ becomes a $(0,1)$-form $\rho_{\bar z}^{J}$, while
$\psi^{J} _{+}$ becomes a scalar  $\chi^{J}$ in the type A model,  and a
$(1,0)$-form $\rho^{J}_{z}$ in the type B model. 

It turns out \cite{tsm} that the type A model can be formulated on any almost
Hermitian  target manifold. However, the type B model was obtained through the
gauging of an  anomalous current, and this can give ill-defined models: the
anomaly in  the global current, given in (\ref{anomalia}), is inherited in the
twisted model as a  global anomaly in the  fermion determinant. This leads to
additional restrictions on the geometry of the  target space, as pointed out
in \cite{mm}, because the fact that the $U(1)_A$ current is  chiral  leads to
a non-linear sigma model anomaly. We will present here a computation of  this 
global anomaly using the strategy of \cite{aetio}.  The fermion kinetic term
of the type B model is: 
\be
S_{\rm B}= \int_{\Sigma} d^2 z{\sqrt h}ig_{{\bar I}J} \{ \chi^{\bar I}D_z 
\rho^{J}_{\bar z} 
+ \theta^{\bar I}D_{\bar z}\rho^{J}_{ z}\},
\label{kinb}
\ee 
where $\theta^{\bar I}=\psi_{+}^{\bar I}$ is an scalar in the twisted theory. 
The effective action is then a section of the line bundle,
\be
{\cal L}={\cal L}_1\otimes {\cal L}_2=({\rm det}\,\ D_z) \otimes 
({\rm det}\,\ D_{\bar z}),
\label{ef}
\ee
and the global, topological anomaly is 
measured by $c_1({\cal L})=c_1({\cal L}_1) + 
c_1({\cal L}_2)$. This can be computed 
using the index theorem for families as in \cite{aetio}. Consider first the 
evaluation map:
\bea
{\hat \phi}: {\rm Map}(\Sigma, M) \times \Sigma &\rightarrow& M \nonumber\\
(\phi, \sigma) &\mapsto& \phi(\sigma).
\label{sigevalu}
\eea
By pulling back  differential forms on $M$ through ${\hat \phi}^*$, we get
differential forms on  ${\rm Map}(\Sigma, M) \times \Sigma$, with the natural
bigrading given by the  product structure: 
\be
{\hat \phi}^*(\omega)={\cal O}^{(0)}_{\omega}+{\cal O}^{(1)}_{\omega}+
{\cal O}^{(2)}_{\omega},
\label{bigrado}
\ee
where the ${\cal O}^{(i)}_{\omega}$ are of degree $i$ with respect to
$\Sigma$. This is precisely the descent procedure of \cite{tsm}. 
The topological obstructions are given by: 
\be
c_1({\cal L}_{1,2})=\int_{\Sigma} 
{\rm ch}({\hat \phi}^*({\overline {TM}})) \{\pm 1 -
{1 \over 2}c_1(\Sigma)\},
\label{cherndet}
\ee
where we only keep the degree two forms on 
${\rm Map}(\Sigma, M)$. In $c_1({\cal L})$ only 
the first descendant of 
${\hat \phi}^*({\rm ch}({\overline {TM}}))$ contributes, and we finally get,
\be
c_1({\cal L})=\int_{\Sigma}c_1(\Sigma){\cal O}^{(0)}_{c_1(M)},
\label{anomala}
\ee
where ${\cal O}^{(0)}_{c_1(M)}={\hat \phi}_{\sigma}^{*}(c_1(M))$ is the 
descendant of zero degree with respect to $\Sigma$,  and ${\hat
\phi}_{\sigma}$ is the map obtained from ${\hat \phi}$ by fixing a point 
$\sigma\in \Sigma$ (the cohomology class of the pulled-back form does not
depend  on the $\sigma$ chosen). This result says then that the anomaly in the
global  current $U(1)_A$ is inherited in the  twisted B model as a sigma model
anomaly. The B model has no topological anomalies  if the target is  a
Calabi-Yau manifold or if the worldsheet is a torus ($c_1(\Sigma)=0$). The
last  possibility is natural,  as in this case  the twist does nothing (the
torus is a hyperk\"ahler manifold) and the original $N=2$  supersymmetric
model is anomaly-free.

\section{Twisting the $U(1)_B$ current in $N=2$ supersymmetric QCD}

To be specific, we will consider $N=2$ supersymmetric  QCD with gauge group
$SU(N)$ and  $N_f=1$ in the fundamental representation (we will follow the
notations  and conventions in \cite{lmpol, tesis}). In this case there is one
$N=2$ matter
 hypermultiplet which contains  a complex scalar isodoublet $q=(q_1,q_2)$,
fermions $\psi_{q \alpha}$, $\psi_{\tilde q \alpha}$, $\overline \psi_{q
\dot\alpha}$, $\overline \psi_{\tilde q  \dot\alpha}$, and a complex scalar
isodoublet auxiliary field $F_i$. The fields $q_i$, $\psi_{q \alpha}$,
$\overline \psi_{\tilde q  \dot\alpha}$, and $F_i$ are in the fundamental
representation of the gauge group, while the fields  $q^{i\dagger}$,
$\psi_{\tilde q \alpha}$, $\overline \psi_{q \dot\alpha}$, and $F^{i\dagger}$
are in the conjugate representation. From the point of view of $N=1$ 
superspace, this multiplet contains two $N=1$ chiral multiplets and therefore
it can be described by two $N=1$ chiral superfields $Q$ and $\widetilde Q$,
{\it i.e.}, these superfields satisfy the constraints $\overline
D_{\dot\alpha} Q=0$ and $\overline D_{\dot\alpha} \widetilde Q=0$. 
While the superfield $Q$ is in the fundamental
representation of the gauge group, the superfield $\widetilde Q$ is in the
corresponding conjugate representation. The component fields of these $N=1$
superfields are:
\bea
Q, \;\; Q^\dagger
 \;\; & \longrightarrow \;\; 
q_1, \;\;   \psi_{ q\alpha}  \;\;  F_2, \;\; q^{1\dagger} \;\; 
\overline\psi_{ q \dot\alpha}, \;\;  F^{2\dagger}, \nonumber\\
\widetilde Q , \;\; \widetilde Q^\dagger 
\;\;& \longrightarrow \;\;  q^{2\dagger}, \;\; \psi_{\tilde q \alpha}, 
\;\; F^{1\dagger}, \;\; q_2, \;\;
\overline \psi_{\tilde q\dot\alpha},   \;\;  F_1.
\label{compas}
\eea
The $SU(2)_I$ current includes now a 
contribution from the bosonic part of the hypermultiplet:
\be
j^{\mu}_a={\overline \lambda}{\sigma}^{\mu}{\sigma_a}
\lambda-iq^{\dagger}\sigma_aD^{\mu}q 
+iD^{\mu}q^{\dagger}\sigma_a q.
\label{corredos}
\ee
The twist of the theory can also be understood as a gauging of the $SU(2)_I$
current,  as it happens with the pure $N=2$ supersymmetric 
Yang-Mills theory. This  is
achieved by  adding to the original Lagrangian  a term $-\omega_{\mu}^a
j^{\mu}_a + q^{\dagger}\omega_{\mu}^a\omega^{\mu b}{\sigma_a}  {\sigma_b}q$.
The kinetic term for the bosons in the resulting theory is then, 
\be
(D_{\mu}+ i\omega_{\mu}^a{\sigma_a})q^{\dagger}
(D_{\mu}-i\omega^{\mu a}{\sigma_a})q,
\label{kinbos}
\ee
where $D_{\mu}$ is the covariant derivative acting on scalars in  the
fundamental of $SU(N)$. We then see that, after the twisting, the bosonic
fields  $q$ become positive-chirality spinors. The fields in  the $N=2$
hypermultiplet are redefined after the twisting as follows: 
\bea
q_i & \longrightarrow& M_\alpha,\nonumber\\
\psi_{q \alpha}  & \longrightarrow& 
-\mu_\alpha /{\sqrt 2}, \nonumber\\
\overline \psi_{\tilde q}^{ \dot\alpha} 
& \longrightarrow&  v^{\dot\alpha}/{\sqrt 2},  \nonumber\\
q^{\dagger i}  & \longrightarrow& 
\overline M^\alpha,   \nonumber\\
\overline \psi_{q  \dot\alpha} 
& \longrightarrow &  \overline v_{\dot\alpha}/{\sqrt 2}, \nonumber\\
\psi_{\tilde q}^ \alpha & \longrightarrow & 
 \overline  \mu^\alpha/{\sqrt 2}.\nonumber\\ 
\label{paloma}
\eea
The motivation for this redefinition is geometrical and can be understood
using the  Mathai-Quillen formulation of the twisted theory \cite{lmna}: the
fields  $\mu_{\alpha}$, $\overline  \mu^\alpha$ are a basis of differential
forms for the  configuration space of monopole fields $M_\alpha$, $\overline
M^\alpha$, and the  fields $v^{\dot\alpha}$, $\overline v_{\dot\alpha}$ are a
basis of fields for the  fibre where the moduli equations take values. 

We have seen that, after the twisting, the bosonic fields  $q$ in the
hypermultiplet become positive-chirality spinors: the gauging of  the
$SU(2)_I$ current makes possible to define $N=2$ supersymmetric 
Yang-Mills theory on a 
curved manifold, but the  obstruction associated to $w_2(X)$ reappears when
matter  hypermultiplets are introduced. From the point of view of 
four-dimensional geometry it would be  desirable to construct twisted $N=2$
supersymmetric QCD on a general four-manifold.  In the case of the 
Seiberg-Witten monopole equations, the issue is precisely to consider  
$\sp^c$-structures, and one would like to extend this possibility to the 
non-abelian generalization of these equations. The  idea is to construct  an
extended twisting procedure by gauging an additional global symmetry of the 
physical theory. As the pure Yang-Mills sector is already well-defined with 
the usual gauging of the
$SU(2)_I$ isospin group, this symmetry can only act  on the matter sector.
Morevover, $\sp^c$-structures involve a $U(1)$ gauge group  associated
to  a line bundle $L$ over the four-manifold $X$. In fact, in four dimensions
we have that,
\be
\sp^c_4=\{(A,B) \in U(2)\times U(2): {\rm det}(A)={\rm det}(B) \},
\label{spincfour}
\ee
and therefore the structure groups of the complex spinor bundles
$S^{\pm}\otimes L^{1/2}$ are $SU(2)_{L,R}\times U(1)$, with the same $U(1)$
action in both  sectors. We should then gauge a global, non-anomalous $U(1)$ 
symmetry in the original $N=2$ theory, acting solely on the matter
hypermultiplets. The  required symmetry is precisely the baryon number. Let us
analyze this in some detail. 

The global anomaly-free symmetry of $N=2$ supersymmetric QCD with $N_f$
hypermultiplets is:
\be
SU(N_f)\times U(1)_{B}\times SU(2)_I.
\label{fsime}
\ee
For $N_f=1$, the baryon number $U(1)_B$ acts on the hypermultiplet as, 
\bea
Q& \rightarrow& {\rm e}^{i\phi}Q, \,\,\,\,\,\,\,\,\ {\widetilde Q} 
\rightarrow {\rm e}^{-i\phi}{\widetilde Q},\nonumber\\
Q^{\dagger} &\rightarrow &{\rm e}^{-i\phi}Q, 
\,\,\,\,\,\,\,\,\ {\widetilde Q}^{\dagger} 
\rightarrow {\rm e}^{i\phi}{\widetilde Q}^{\dagger}.
\label{barion}
\eea
As it is a vector symmetry, it is non-anomalous. In components it reads:
\bea  
q& \rightarrow& {\rm e}^{i\phi}q, \,\,\,\,\,\,\,\,\ q^{\dagger} 
\rightarrow {\rm e}^{-i\phi}q^{\dagger}, \nonumber\\
\psi_{q \alpha}&\rightarrow &{\rm e}^{i\phi}\psi_{q \alpha}, 
\,\,\,\,\,\,\,\,\ 
{\overline \psi}_{{\tilde q} \dot  \alpha} 
\rightarrow {\rm e}^{i\phi}
{\overline \psi}_{{\tilde q} \dot  \alpha},\nonumber\\
{\overline \psi}_{ q \dot \alpha}&\rightarrow &{\rm e}^{-i\phi}
{\overline \psi}_{ q \dot \alpha}, 
\,\,\,\,\,\,\,\,\ 
\psi_{ {\tilde q} \alpha}\rightarrow 
{\rm e}^{-i\phi}\psi_{{\tilde q} \alpha}.
\label{barioncom}
\eea
The $U(1)_B$ current associated to this symmetry is:
\be
j^{\mu}_{B}=-iD_{\mu}q^{\dagger}q+iq^{\dagger}D^{\mu}q+ 
{\overline  \psi}_{{ q} \dot  \alpha}
(\sigma^{\mu})^{\dot \alpha \alpha}{\psi}_{{ q} \dot  \alpha}-
{\overline  \psi}_{{\tilde q} \dot  \alpha}
(\sigma^{\mu})^{\dot \alpha \alpha}
{\psi}_{{\tilde q} \dot  \alpha}.
\label{barcor}
\ee
To gauge this $U(1)$ symmetry, consider the determinant line bundle $L$
associated to a $\sp^c$-structure on $X$, endowed with a connection
$b_{\mu}$, and add to the  Lagrangian the term: 
\be
{1 \over 2}j^{\mu}_B b_{\mu} -{1 \over 4}q^{\dagger}b_{\mu}b^{\mu}q.
\label{copla}
\ee
If we gauge both the $SU(2)_I$ and the $U(1)_B$ symmetries, the covariant
derivatives  acting on the components  of the matter hypermultiplet in the
resulting Lagrangian are the appropriate ones  for complex spinors taking
values  in $S^{\pm}\otimes L^{1/2}\otimes E $. To further analyze the
consistency of the  procedure, it is  useful to consider the correspondence
between the fields in the original $N=2$ theory  and the fields appearing in
the Mathai-Quillen formulation of the moduli problem.  This correspondence is
given in (\ref{paloma}), and the structure of (\ref{copla}) implies 
that the components of the matter hypermultiplet will be sections of 
$L^{\pm {1\over 2}}$ if their baryon number is $\pm 1$.
 Taking into account the baryon number  assignment in
(\ref{barioncom}), we see that the  fields $M_{\alpha}$, $\mu_{\alpha}$ are
sections of $S^{+}\otimes L^{1/2} \otimes E$,  $v^{\dot \alpha}$ is a section
of $S^{-}\otimes L^{1/2}\otimes E$, ${\overline M}^{\alpha}$,  ${\bar
\mu}^{\alpha}$ are sections of $S^{+}\otimes L^{-1/2}\otimes {\overline E} $,
and  ${\bar v}_{\dot \alpha}$ is a section of $S^{-} \otimes L^{-1/2}\otimes
{\overline E}$.  This is then  consistent with the expected structure of the
complex spinor bundles, with the  same determinant line bundle for both
$S^{\pm}$. The  kinetic terms in the twisted theory are then, 
\be
{\cal L}_{\rm k}=g_{\mu \nu}D^{\mu}_{L}{\overline
M}^{\alpha}D^{\nu}_{L}M_{\alpha}-  { i \over 2}\Big({\bar v}_{\dot
\alpha}D_{L}^{\dot \alpha \alpha}\mu_{\alpha}+ {\bar \mu}^{\alpha}D_{L \alpha
\dot \alpha}v^{\dot \alpha} \Big),
\label{kinbardos}
\ee
where,
\be
D^{\mu}_{L}=D_{\mu}+{i\over 2}b_{\mu},
\label{spincov}
\ee
is the covariant derivative associated to the tensor product connection on 
$S^{+} \otimes L^{1/2}\otimes E $, $D_{\mu}$ is given in (\ref{cova}), 
and $D_L^{\dot \alpha \alpha}$ is the 
corresponding Dirac operator. The remaining terms in the Lagrangian are the 
same as in the non-abelian monopole theory constructed in \cite{lmna}.

As in the case of the usual twisting involving matter  hypermultiplets
\cite{lmt, tesis}, the Lagrangian obtained after  the gauging of the 
$SU(2)_I$ and the $U(1)_B$ symmetries is not $Q$-closed. This is a geneal
feature of  the twisting procedure: in order to guarantee the existence of a
scalar topological  symmetry on a curved manifold it can be necessary to add to
the Lagrangian  terms involving the non-trivial geometry of this manifold.
 If we compute $[Q,{\cal L}]$,  we must use the Weitzenb\"ock formula  for the
$\sp^c$-case \cite{sping}: 
\be
D^{\dagger}_{L\alpha\dot\alpha}
D^{\dot\alpha \beta}_L = (D_{L\mu} D^\mu_L  + 
{1\over 4} R)\delta_{\alpha}{}^{\beta}
+i F_{\alpha}{}^{\beta}+ {i \over 2}\Omega_{\alpha}{}^{\beta}, 
\label{miwei}
\ee 
where $R$ is the scalar curvature of the manifold, and
$F_{\alpha}{}^{\beta}$  and $\Omega_{\alpha}{}^{\beta}$ are the self-dual
part  of the gauge field strength on $E$ and the curvature of the 
line bundle $L$, respectively. The kinetic terms in  (\ref{kinbardos}),
${\cal L}_{\rm k}$,  give a non-zero contribution:  
\be
[ Q,{\cal L}_{\rm k} ]=-{R \over 4}({\overline M}^{\alpha}
\mu_{\alpha}+{\bar \mu}^{\alpha} 
M_{\alpha})-
{i \over 2}\big( {\overline M}^{\alpha}
\Omega_{\alpha}{}^{\beta}\mu_{\beta}+
{\bar \mu}^{\alpha}\Omega_{\alpha}{}^{\beta} M_{\alpha}\big).
\label{noexdos}
\ee
It then follows that the modified Lagrangian,
\be
{\cal L}_{\rm top}={\cal L}+{R \over 4}
{\overline M}^{\alpha}M_{\alpha}+{i \over 2}
{\overline M}^{\alpha}\Omega_{\alpha}{}^{\beta}M_{\alpha},
\label{totexdos}
\ee
where ${\cal L}$ contains the kinetic term 
${\cal L}_{\rm k}$ in (\ref{kinbardos}) plus the rest of the terms
of the theory of non-abelian monopoles \cite{lmna},
is $Q$-closed on a general
four-manifold. This Lagrangian was obtained in 
\cite{tqcd}, and a standard analysis shows that the resulting 
topological field theory corresponds 
to the moduli problem encoded in the equations
\bea
F_{\alpha\beta}^{+a}+{i}{\overline
M}_{(\alpha} (T^{a}) M_{\beta )}&=&0,\nonumber\\
D_{L}^{\alpha\dot\alpha}M_\alpha &=&0.
\label{compactoc}
\eea
These are equations for a pair $(A,M)$ consisting of a connection  $A$ on $E$
and a section $M$ of the complex spinor bundle $S^+\otimes L^{1/2}  \otimes
E$, where the connection on the determinant line bundle $L$ is fixed. The
operator  $D_{L}$ is just the Dirac operator for this twisted bundle. Similar
equations  have been considered in the mathematical literature, see 
\cite{okone,bradgp,gptres,pt}. Notice that the twist of the $U(1)_B$
current  gives precisely the geometrical content of these moduli equations. 
As it was expected, the resulting topological field theory is anomaly free, 
because the determinat line bundle of the twisted Dirac operator $D_L$ 
is always orientabledue to its underlying complex structure.

Usually, the fact that the theory is topological means that correlation
functions  do not depend on the Riemannian metric of the four-manifold. In the
same way one can  easily check that the theory is topological with respect to
the $\sp^c$-connection:  the correlation functions do not depend on the
choice of the connection $b_{\mu}$  on the line  bundle $L$, but only on the
topological class of the $\sp^c$-structure. To see this,  notice that
the Mathai-Quillen formulation of the model coupled to a $\sp^c$-structure is 
almost identical to the one presented in \cite{lmna}, the only difference 
being that we must consider instead the  expression for the Dirac operator
$D_L$ (including the connection $b_{\mu}$ on the  determinant line bundle
$L$). The $Q$-transformations of  the fields in this off-shell formulation are
identical to the ones in the $\sp$ case
 and do not depend on  the connection on $L$. As the full Lagrangian
(\ref{totexdos}) is  $Q$-exact, we have, using the results in \cite{lmna}: 
\be
{\delta {\cal L}_{\rm top} \over \delta b_{\mu}}=
\{Q, {1 \over 2}({\bar v}_{\dot \alpha} 
(\sigma^{\mu})^{\dot \alpha \alpha}M_{\alpha}-{\overline M}^{\alpha}
(\sigma^{\mu})_{\dot \alpha \alpha}v^{\dot \alpha}) \}.
\label{topscon}
\ee
This in turn guarantees that the twisted theory is independent of the 
choice of $b_{\mu}$. 

Notice that the metric (or Spin connection) and the $\sp^c$-connection
enter  the construction on the same footing. In this formulation of
non-abelian monopoles  coupled to $\sp^c$-structures, the connection on the
determinant line bundle $L$ is a  background gauge field just like the $\sp$
connection. The analysis of the moduli problem  associated to the equations
(\ref{compactoc}) is very  similar to the one presented in \cite{lmna} in the
$\sp$ case, but one should not  divide by the group of gauge transformations
associated to the $U(1)$ gauge field,  because this is a {\it background}
field. This is in contrast with the corresponding  situation in the abelian
theory. However, the topological correlation functions  depend on the
topological  class of the $\sp^c$-structure chosen to gauge the  $U(1)_B$
symmetry. In particular, the virtual dimension of the  moduli space depends
now on the first Chern class of $L$, and other features (like the 
orientability, the analysis of reducibles and the structure of the
observables) are almost  identical to the $\sp$ case.   

\section{Conclusions and outlook}
We have shown that $N=2$ supersymmetric QCD has the possibility of coupling the
matter fields to  $\sp^c$-structures once the adequate symmetry has been
identified. The gauging of the  $U(1)_B$ current can be generalized to
theories with more than one hypermultiplet: in this  case there are $N_f$
$U(1)$ symmetries that can be gauged, and this makes possible to  consider
$N_f$ different $\sp^c$-structures, as it has been already noticed in
\cite{tqcd}.  An obvious extension of this work would be the computation,
using physical methods, of  the topological correlation functions of this
extended model, generalizing in this way  the results in \cite{lmpol}. One
should be careful with the subtleties involving  $\sp^c$-structures that have
been discussed in \cite{abelia}. But the moral of our
 procedure  is perhaps that in the construction of  topological quantum field
theories one could consider not only
the  symmetries associated to the supersymmetry algebra, but {\it any} global
symmetry of the theory which might have a geometrical meaning through an
appropriate gauging.  
 
\begin{center}
\large{ACKNOWLEDGEMENTS}
\end{center}
We would like to thank L. \'Alvarez-Gaum\'e and O. Garc\'\i a-Prada 
for useful discussions. 
This work was supported in part by DGCIYT under grant PB93-0344.

\end{document}